# Comparison of Uncertainty of Two Precipitation Prediction Models

Stephen Shield[1,2] and Zhenxue Dai[1]

[1] *Earth & Environmental Sciences Division, Los Alamos National Laboratory Los Alamos, NM-87545*

[2] *Department of Earth and Atmospheric Sciences, University of Nebraska-Lincoln, Lincoln, NE 68588*

**Abstract:** Meteorological inputs are an important part of subsurface flow and transport modeling. The choice of source for meteorological data used as inputs has significant impacts on the results of subsurface flow and transport studies. One method to obtain the meteorological data required for flow and transport studies is the use of weather generating models. This paper compares the difference in performance of two weather generating models at Technical Area 54 of Los Alamos National Lab. Technical Area 54 is contains several waste pits for low-level radioactive waste and is the site for subsurface flow and transport studies. This makes the comparison of the performance of the two weather generators at this site particularly valuable.

## Introduction

Technical Area 54(TA-54) at Los Alamos National Lab contains several waste pits for disposal of low-level waste, certain radioactively-contaminated infectious waste, asbestos-contaminated material, and polychlorinated biphenyls (Los Alamos National Lab, 2015). It is important to ensure subsurface flow and transport of waste does not pose a risk to the environment in the future. Specifically, there is some risk of the waste eventually being transported into the ground water which lies several hundred meters below the surface (Dai et al., 2010; 2012). Model simulations are conducted to predict the water flow below the pits; however, they require the input of future meteorological data to provide subsurface flow predictions (Domenico and Schwartz, 1990).

Weather generator models were chosen as a method to generate the 1,000 year long meteorological data sets needed for input into sub-surface flow models. Weather generators such as WeaGETS are capable of generating long time series predictions of precipitation, which occasionally result in extremes much larger than those recorded at weather stations (Caron et al., 2008). For example, a major precipitation event with a probability of happening once in 1,000 years is unlikely to have happened during the short period of observed data available. However, in a study predicting the next 1,000 years, it is likely for such an event to happen during that time. Due to the significance of such extreme events in a meteorological data set it is important to include such events (Duan et al., 1992).

This paper compares the uncertainty in the quality of performance for two different weather variables of two weather generator predictions for TA-54 at Los Alamos National Lab.

## Data and Methods

WGEN is a weather generator originally created by C. W. Richardson in 1981 (Williams et al., 2008). Since then it has become well know and has been widely used. It has also been incorporated into a number of applications and other programs (Soltani et al., 2000). Several more recently developed weather generators provide advantages over the more popular generator WGEN and for this study they were chosen over WGEN.

**WeaGETS:** WeaGETS is a daily stochastic weather generator that can generate precipitation, maximum temperature, and minimum temperature time series of unlimited length for use in agricultural and hydrological impact studies (Chen et al., 2012). WeaGETS was selected as one of the weather generators due to its superiority over the weather generator WGEN (Caron et al., 2008). WeaGETS has the ability to better predict long wet and dry spells than WGEN because it has the ability to use a higher-order Markov model while WGEN uses a first-order two-state Markov model (Bastola et al., 2011).

The following is a brief summary of what user controls are available in WeaGETS and which option was chosen for the generation of the data. For more information on WeaGETS see the user manual which was used to determine the best user options for this project

The WeaGETS weather generator has several user controllable smoothing schemes for precipitation. This can be useful as the model generates weather for a time scale of two weeks and without smoothing there can be sudden changes in weather patterns not present in the observed data. The use of a Fourier harmonic reduces these sudden changes but when higher-order Fourier harmonics are used it can also result in weather patterns that are not present in the observed data. A third-order Fourier harmonic was chosen to attempt to balance between these two possible outcomes. Choosing a higher order Fourier harmonic should result in a better output due to the local climate of the Los Alamos area having distinct rainy and dry seasons.

WeaGETS also has options for different order Markov models to generate precipitation frequency. The higher-order models are better at generating long dry or wet spells. This makes a higher-order Markov model advantageous due to Los Alamos's climate. For precipitation frequency a third-order Markov model was used to predict frequency of precipitation. This was done with the knowledge that higher-end Markov models have some risk of error when used with a relatively short period of input data. However, it is believed the benefit of better precipitation prediction outweighs the risk of error.

Additionally WeaGETS offers two options to compute the precipitation amount when smoothing is selected. For this study the gamma distribution was chosen due to its simplicity and superiority to the exponential distribution.

WeaGETS also requires a minimum precipitation threshold to determine if a given day will be categorized as a wet day or a dry day. Typically 0.1mm is selected as this threshold.

However, since the minimum precipitation threshold in the observed data was 0.01 inches or 0.254mm that value was used instead.

For temperature generation WeaGETS offers two schemes: an unconditional scheme where the maximum and minimum temperature are calculated separately and a conditional scheme where the minimum temperature is calculated based off of the days maximum temperature. The conditional method was selected because it was more likely to correctly predict diurnal changes in temperature.

**LARS-WG:** LARS was chosen as the other weather generator due to its ability of reproducing observed interannual variability of rainfall frequency (Mavromatis and Hansen, 2001). It also has the ability to reproduce the means of yearly maxima for daily precipitation and 10 and 20 year return values for precipitation accurately (Semenov, 2008). LARS also has some advantages over WGEN. LARS uses semi-empirical distributions which are more flexible than the standard distributions used by WGEN. This allows LARS to perform well in a range of diverse climates (Semenov, 1998).

LARS has less user-controls than WeaGETS; however, one option it does have is its ability to predict future daily weather under different climate change scenarios. It is important to note that this feature was not used and the generated weather data should be statistically similar to the observed data set for the entire time period.

**Data:** Weather generators require the input of observed data in order to generate a future meteorological data set. For this required data observed data daily weather data for the time period January 29, 1992, to July 12, 2015, was used. Data was collected at TA-54 of Los Alamos National Lab and was obtained through the LANL Weather Machine. Weather variables obtained for input into weather generators include daily high and low temperatures, daily total precipitation, and daily total solar radiation. Data that was missing or suspect was removed prior to retrieval and coded as required by each weather generator prior to input.

The weather generators WeaGETS and LARS-WG were then ran to generate 1,000 year long meteorological data sets which could be used in subsurface flow and transport models. A summary of the two weather generators and the options selected is located below.

**Summary:** In order to analyze the large amount of data generated, monthly summaries of daily data were created for each month in 1,000 year long time series, as well as for the observed data. Monthly average daily high and low average temperatures were calculated as well monthly total precipitation. This was done using a Visual Basic for Applications code within Microsoft Excel and yielded a total of 12,000 data points for each meteorological variable for both weather generator models as well as a smaller data set of the observed data.

This data was then analyzed using R, a programing language and software environment for statistical computing and graphics, to determine the monthly distributions for each variable. Based on the results of the distributions for each variable, an appropriate statistical comparison method was chosen to determine which of two weather generating models produced data most similar to the observed data.

Plots were then generated to show how each generator performed for each variable on a monthly basis. These plots also show the overall performance of each weather generating model.

## Analysis and Discussion

**Maximum Temperature:** The distributions of daily temperature are shown in Figure 1. The observed monthly average daily maximum temperature data shows a majority of monthly distributions are bimodal with several months exhibiting fat-tailed or skewed distributions. The data generated by WeaGETS shows greater variety in the distributions with several months each of bimodal, fat-tailed, and normal or nearly normal distributions. The LARS data shows most months with normal to nearly normal distributions with a few months exhibiting a very small amount of skewness.

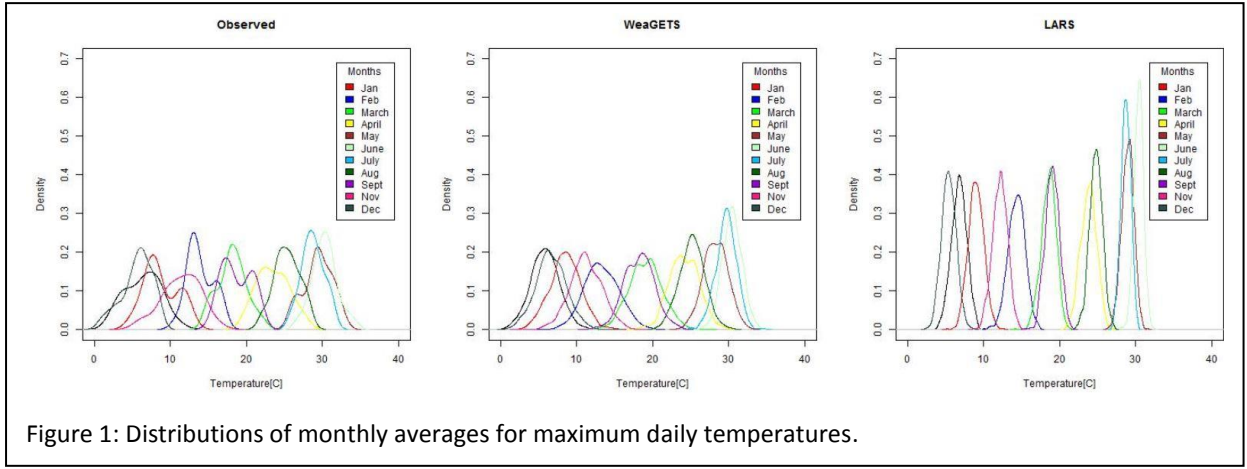

Figure 1: Distributions of monthly averages for maximum daily temperatures.

Due to the large number of non-normal distributions, especially in the observed data, analyzing performance of the weather generators using the comparison of means method would likely result in an inaccurate assessment. This is due to the mean being a poor representation of non-normal distributions. The median is a much better representation of non-normal data distributions (McCluskey and Lalkhen, 2007). For this reason comparison of medians was used for analyzing performance of the weather generators in generating daily maximum temperatures.

Figure 2 shows the magnitude of the difference from the median of the observed data set to the medians of the generated data sets. The results show that WeaGETS outperforms LARS six months of the year while LARS outperforms WeaGETS during the other six months. WeaGETS performs well during the late winter and early spring and again in the fall. LARS preforms better during most of the spring and summer. Taking the average of monthly differences for the entire year we see both generators have less than 0.6 Degrees Celsius difference in medians. LARS does have a slight advantage overall with 0.027 Degrees Celsius less difference between average medians.

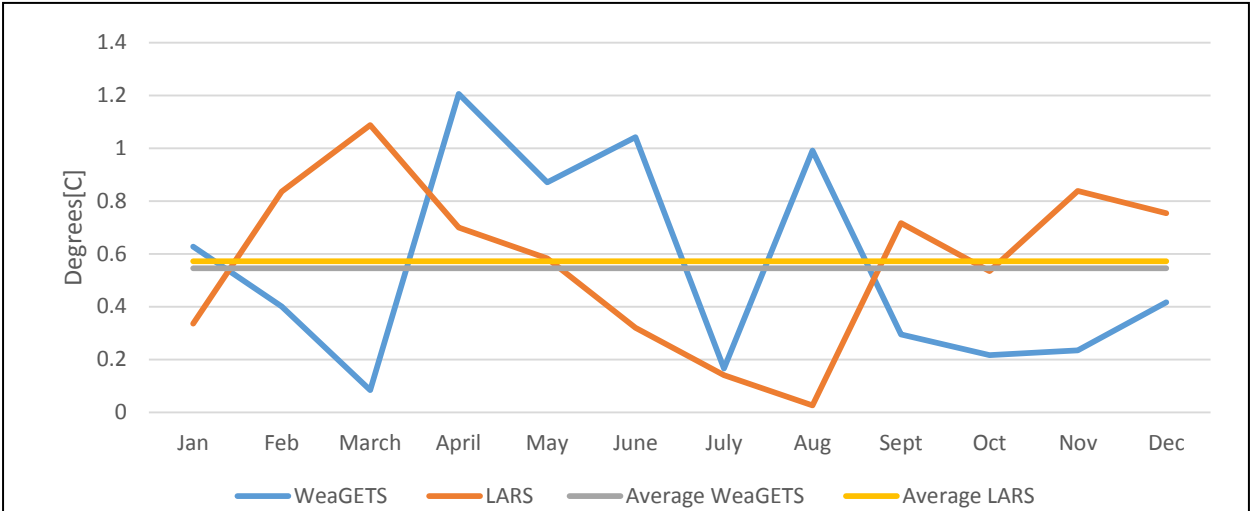

Figure 2: Magnitude of difference from observed median daily maximum temperature and generated median daily maximum temperature.

**Minimum Temperature:** The distributions of monthly averages of daily minimum temperature are shown in figure 3.

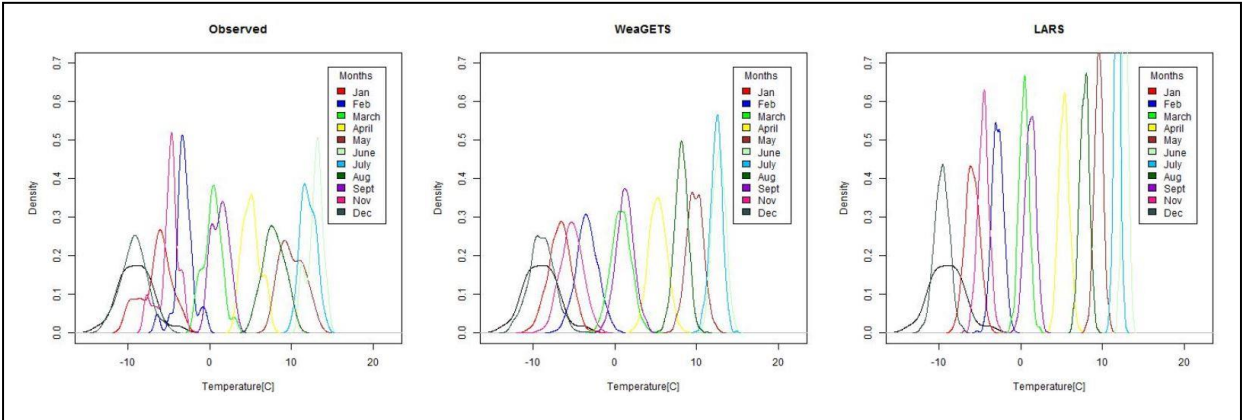

Figure 3: Distributions of monthly averages for minimum daily temperatures.

The observed data once again show a number of months with bimodal and fat-tailed distributions. A few months also show near-normal distributions. The WeaGETS data set shows a larger number of months with near-normal distributions. Several months still exhibit a slight skewness or slight bimodal characteristics. LARS once again shows a large number of months with near-normal distributions. However, several months still exhibit a slight skewness.

Once again, a number of non-normal distributions exist especially in the observed data. Therefore, the method of comparison of medians again used to analyze performance of generating daily minimum temperature.

Figure 4 shows the magnitude of the difference from the median of the observed data set to the medians of the generated data sets. For daily minimum temperature the LARS weather generator shows superiority for 8 out of 12 months. There is no clear season where the WeaGETS has an advantage. The months it preforms better are scattered throughout the year. Overall, both generators had average difference in medians for minimum temperatures less than the difference in medians for the high temperatures. This means both generators are more accurate when generating daily minimum temperatures than they when generating daily high temperatures. For minimum temperature LARS once again had the advantage, but by a larger margin, with a monthly average difference in median 0.123 Degrees Celsius less than WeaGETS.

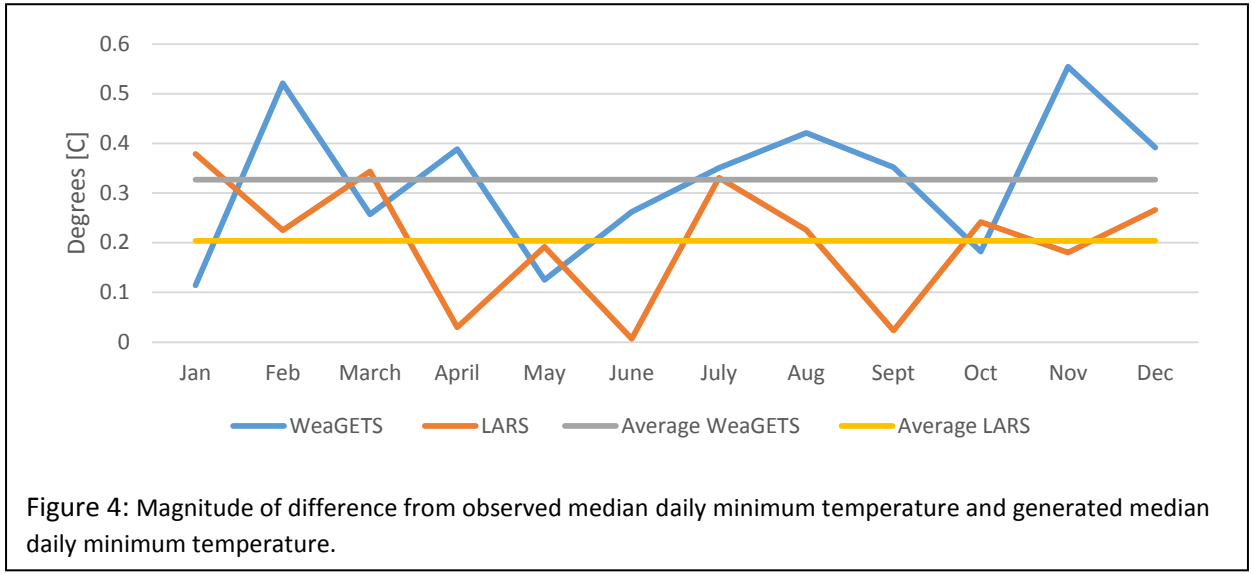

Figure 4: Magnitude of difference from observed median daily minimum temperature and generated median daily minimum temperature.

**Precipitation:** Figure 5 shows the monthly distributions for monthly total precipitation.

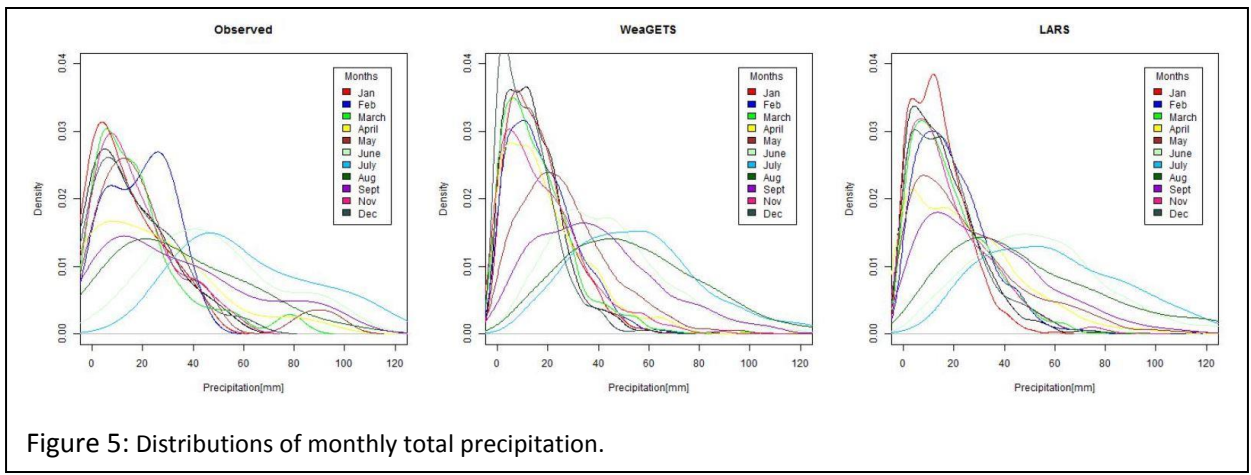

Figure 5: Distributions of monthly total precipitation.

As would be expected for precipitation the distributions are heavily right skewed due to the large number of days in which no precipitation falls. During the summer monsoon season the distributions are fat-tailed due to the more frequent days with precipitation as well as the presence of relatively infrequent but heavy precipitation events. These characteristics are seen for both weather generators as well as the observed data set. Due to the almost exclusive non-normal distributions we will once again utilize the method of comparing medians to judge weather generator performance.

Figure 6 shows the magnitude of the difference from the median of the observed data set to the medians of the generated data sets.

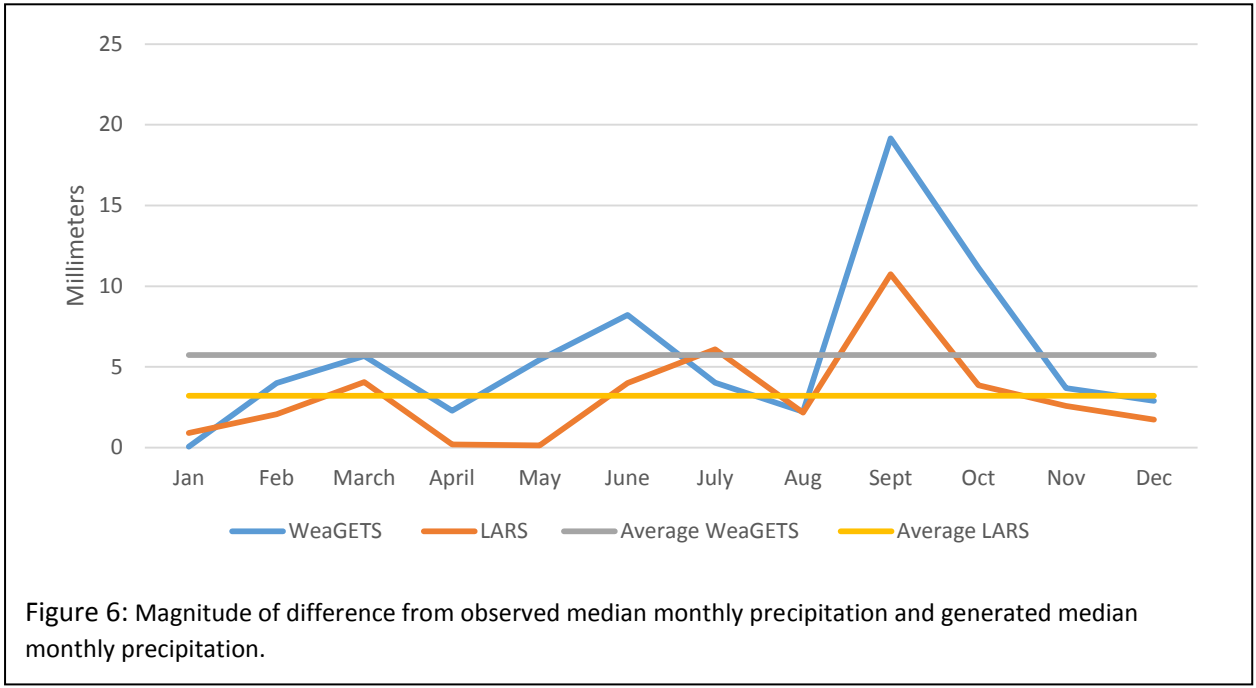

Figure 6: Magnitude of difference from observed median monthly precipitation and generated median monthly precipitation.

With the exceptions of January and July, LARS outperformed WeaGETS. LARS' average difference in median was 2.523 mm less than WeaGETS' average difference in median. This indicates that overall LARS produced distributions closer to the observed distributions than WeaGETS.

Because precipitation is the most significant meteorological variable when considering subsurface flow and transport, further analyzation may be beneficial for decision-making. Figure 7 shows both the magnitude of the difference in medians and whether the difference was positive or negative.

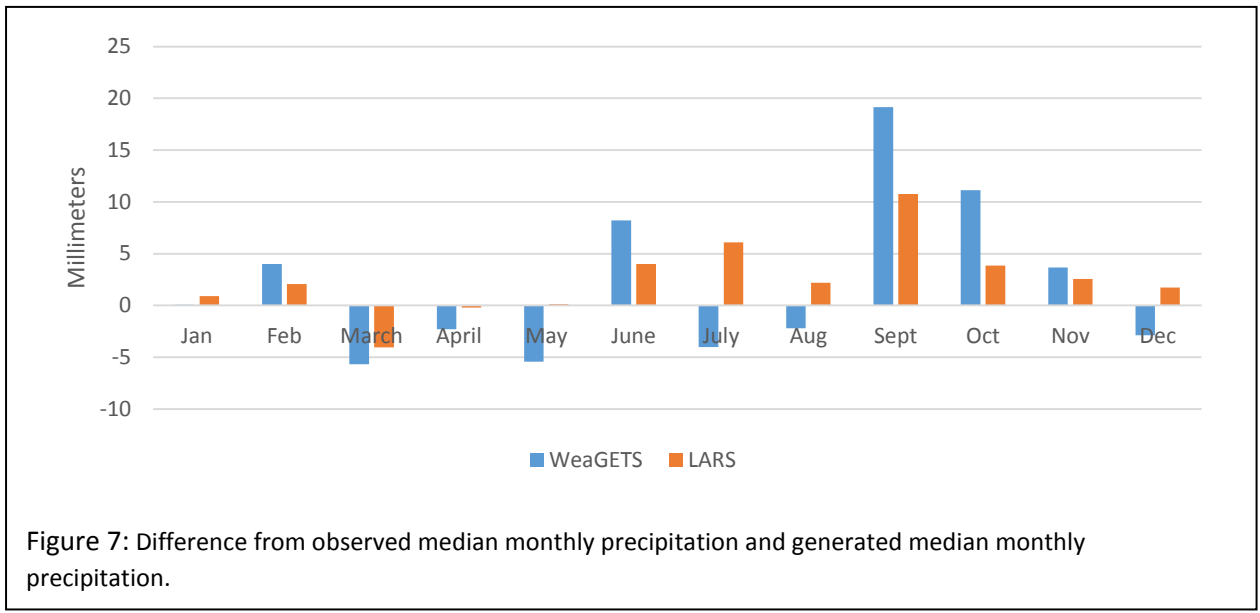

Figure 7: Difference from observed median monthly precipitation and generated median monthly precipitation.

    WeaGETS shows a slight bias for under-estimation of precipitation during the dry season with under-estimation occurring during March, April, and May. The over-estimation during the transition month of June followed by under estimation during July and August suggests that while WeaGETS is able to quickly pick up on the start of the rainy season, it does poorly in reproducing the large increase in monthly precipitation during the rainy season. The over-estimation that occurs during the slow transition back to the dry season also suggests WeaGETS is slow to adjust back to the dry season.

    LARS doesn't show a clear bias for over or underestimation during the dry season with over estimation occurring some months and underestimation occurring in other months. LARS also quickly picks up on the transition to the rainy season over estimating precipitation in June. LARS continues to overestimate precipitation during both rainy summer season and fall transition months. This consistent over-estimation during the rainy season may cause errors in subsurface flow models due to the long periods of rainfall above observed rainfall being present in the meteorological data set.

    Both weather generators show significant over estimation of precipitation during the month of September. One possible cause for this is the historic rainfall that occurred during mid-September of 2013. (NWS Albuquerque, 2013) During this time period 172.72 mm (6.8 in.) of precipitation fell at TA-54. Of this precipitation, 108.712mm (4.28in.) fell during a 48 hour period. It appears both weather generators had trouble judging the rarity of this event and over produced such events in their output. The likely reason LARS over produced this type of event is the semi-empirical distribution used for generation of precipitation. The flexibility of the distributions allows for extreme events to be modeled directly (Semenov et al., 1998). WeaGETS also does a poor job of producing extreme precipitation events. This is likely because many of these largest extreme values are associated with unusual meteorological events (e.g., hurricanes or mesoscale convective complexes) which suggests these extreme precipitation events come from different populations than most of the daily precipitation observations to which the distributions have been fit (Wilks, 1999). The use of a mixed distribution scheme may improve performance for extreme events (Li, 2012). This because mixed distribution schemes contain two

distributions; one for small rainfall events and one for large rainfall events. Unfortunately, at this time neither weather generator offers a mixed distribution scheme for precipitation generation.

While the results of this study are important to consider for studies at TA-54, further research should be conducted on the performance of the weather generators before these observations should be considered proven tendencies that will occur at other sites or with a longer input data set.

## Summary and Conclusion

Overall, the LARS weather generator generated data are more similar to the observed data than WeaGETS for all three meteorological variables. This would make it a better choice for input into subsurface flow and transport models and as its use is more likely to result in accurate predictions of subsurface flow and transport of waste material at TA-54. That said, LARS is not without flaws. In particular the consistent over estimation of precipitation during the rainy season is a cause for concern as it could be a source of error in subsurface flow and transport models. It also appears that both LARS and WeaGETS had difficulty handling extreme precipitation events. The data does suggest LARS handled this type of extreme event better than WeaGETS with less difference in median precipitation for the month of September. The use of a longer time series of input data would likely help reduce this source of uncertainty and would likely improve the overall quality of the output data as well as helping the weather generators handle extreme events.